\documentclass[showpacs,aps,preprint]{revtex4}
\usepackage{graphicx}
\usepackage{amsmath,amssymb}
\usepackage{accents}
\makeatletter
\def\widebar{\accentset{{\cc@style\underline{\mskip10mu}}}}
\makeatother

\def\gtsim{\mathrel{\hbox{\raise0.2ex
  \hbox{$>$}\kern-0.75em\raise-0.9ex\hbox{$\sim$}}}}
\def\ltsim{\mathrel{\hbox{\raise0.2ex
  \hbox{$<$}\kern-0.75em\raise-0.9ex\hbox{$\sim$}}}}

\begin{document}

\title{
Rotational motion of triaxially deformed nuclei studied by
microscopic angular-momentum-projection method II: \\
Chiral doublet band
}

\author{Mitsuhiro Shimada, Yudai Fujioka, Shingo Tagami and Yoshifumi R. Shimizu}
\affiliation{Department of Physics, Graduate School of Science,
Kyushu University, Fukuoka 819-0395, Japan}


\begin{abstract}

In the sequel of the present study, we have investigated
the rotational motion of triaxially deformed nucleus
by using the microscopic framework of angular-momentum projection.
The Woods-Saxon potential and the schematic separable-type interaction
are employed as a microscopic Hamiltonian.
As the first example nuclear wobbling motion
was studied in detail in the part~I of the series.
This second part reports on another interesting rotational mode,
chiral doublet bands:
two prototype examples, $^{128}$Cs and $^{104}$Rh, are investigated.
It is demonstrated that the doublet bands naturally appear
as a result of the calculation in this fully microscopic framework
without any kind of core,
and they have the characteristic properties
of the $B(E2)$ and $B(M1)$ transition probabilities,
which are expected from the phenomenological
triaxial particle-rotor coupling model.

\end{abstract}

\pacs{21.10.Re, 21.60.Ev, 23.20.Lv}

\maketitle

\section{Introduction}
\label{sec:intro}

It has been an interesting subject
to study the rotational motion of triaxially deformed nucleus
in the field of nuclear structure~\cite{BM75}.
Although the triaxial deformation rarely realized
in the ground state of nuclei~\cite{Mol06},
it is more frequently expected at high-spin states,
see, e.g., Refs.~\cite{VDS83,Fra01,Pan11} and references therein.
When the nuclear mean-field is triaxially deformed,
collective rotation about all three principal axes is possible,
and therefore the total angular-momentum vector may tilts away
from either of three principal axes.
Then the quantized rotational spectrum of the rigid-rotor will emerge,
which is called nuclear wobbling~\cite{BM75}.
Such rotational bands have been sought for a long time and finally
identified first in $^{163}$Lu~\cite{Ode01};
see, e.g., Refs.~\cite{NMM16,Fra17} for recent theoretical review articles.
This wobbling motion has been investigated
in the first part of the present study~\cite{SFTSI};
we refer to it as part~I hereafter.

Another specific rotational motion expected in a triaxially deformed nucleus
are the chiral doublet bands, predicted for the first time in Ref.~\cite{FM97},
where the tilting of the angular-momentum vector
is caused by other degrees of freedom than the collective rotation;
see e.g., Ref.~\cite{SK17} for a recent review.
For an odd-odd nucleus, where both an odd proton particle and
an odd neutron hole occupies a high-$j$ intruder orbitals,
as a typical example,
the odd particle angular-momentum aligns along the short axis and
the odd hole angular-momentum along the long axis,
because such alignments maximize the overlap of the wave function
of the aligned particle or hole
with triaxial density distribution of the core.
If the three moments of inertia of the core are in irrotational-like ordering,
the collective angular-momentum aligns along the medium axis,
which has the largest moment of inertia.
For moderate high-spin states,
where all three kinds of angular momenta are sizable,
these three vectors are aplanar, and the chiral symmetry
between the right- and left-handedness is broken in such a system.
It is then expected that
a pair of degenerate ${\mit\Delta}I=1$ rotational bands
appear as a result of breaking this symmetry.
In Ref.~\cite{KSH04}, it is discussed that
characteristic patterns are expected
for the electromagnetic transition rates, $B(E2)$ and $B(M1)$
in this prototype situation with broken chiral symmetry.

These interesting types of rotational motion
characteristic for triaxially deformed nucleus have been investigated
mainly by phenomenological models such as
the triaxial-rotor~\cite{BM75}
or the particle-hole coupled to triaxial-rotor~\cite{FM97}.
Here we study such rotational motion by
employing the fully microscopic framework,
where the nuclear wave function is constructed from the triaxially
deformed mean-field and the broken rotational-symmetry
is recovered by angular-momentum projection;
see, e.g., Ref.~\cite{RS80}.  With the projection method,
the regular rotational spectrum is naturally obtained.
Full 3D projection from the mean-field wave function
should be performed for triaxially deformed nuclei,
so that an efficient method is necessary.
We have developed such a method in Ref.~\cite{TS12},
and applied it to the study of
nuclear tetrahedral deformation~\cite{TSD13,TSD15},
the $\gamma$-vibration~\cite{TS16},
and the ground-state rotational bands~\cite{STS15,STS16} in rare earth nuclei.
In this second part of the present investigation
we also employ the same method to study the chiral doublet band
for the case where the prototype considered in Ref.~\cite{KSH04}
is realized.

It should be mentioned that chiral doublet bands have been studied
by a similar microscopic approach, the triaxial projected shell model,
for the first time in Ref.~\cite{BSP12};
see Refs.~\cite{Sun16,SBD16} for recent review articles.
The authors are successful to reproduce the experimental data.
The purpose of the present work is not to reproduce the experimental data,
but rather to understand how the chiral doublet bands appear and
how the ideal chiral geometry reflects to the observable quantities
such as the electromagnetic transition rates.
We believe that such an investigation is meaningful for deeper comprehension
of the rotational motion in the triaxially deformed nucleus
from the microscopic view point.

The paper is organized as follows.  We briefly recapitulate our formulation
in Sec.~\ref{sec:formulation}, where only the necessary mathematical
expressions for discussion of the present study are included.
The more detailed content is presented in part~I~\cite{SFTSI}.
Possible occurrence of the chiral doublet band and
its properties of the electromagnetic transition probabilities
are studied for $^{128}$Cs and $^{104}$Rh nuclei in Sec.~\ref{sec:chiral}.
Finally the results of the present study are summarized
in Sec.~\ref{sec:summary}.
Preliminary results were already published in Ref.~\cite{TSF14}.

\section{Basic Formulation}
\label{sec:formulation}

In the series of the present work, we study
collective rotation of triaxially deformed nucleus
with the microscopic angular-momentum-projection method.
The quantum eigenstates of rotational band are obtained by
\begin{equation}
 |\Psi_{M\alpha}^{I}\rangle =
 \sum_{K} g_{K,\alpha}^{I}\,
 \hat P_{MK}^I|\Phi \rangle
\label{eq:wfProj}
\end{equation}
from the mean-field state $|\Phi \rangle$,
where the angular-momentum projector is denoted by $\hat P_{MK}^I$
and the amplitude $g^I_{K,\alpha}$ is determined by
the so-called Hill-Wheeler equation, see, e.g., Ref~\cite{RS80};
\begin{equation}
 \sum_{K'}{\cal H}^I_{K,K'}\ g^I_{K',\alpha} =
 E^I_\alpha\,
 \sum_{K'}{\cal N}^I_{K,K'}\ g^I_{K',\alpha},
\label{eq:HW}
\end{equation}
with the definition of the Hamiltonian and norm kernels,
\begin{equation}
 \left\{ \begin{array}{c}
 {\cal H}^I_{K,K'} \\ {\cal N}^I_{K,K'} \end{array}
 \right\} = \langle \Phi |
 \left\{ \begin{array}{c} \hat H \\ 1 \end{array}
 \right\} \hat{P}_{KK'}^I | \Phi \rangle.
\label{eq:kernels}
\end{equation}
To investigate how the interesting types of rotational motion appear
and what kind of properties they have,
it is preferable to be able to change the mean-field parameters,
e.g., the deformation parameters, arbitrarily.
Therefore, we employ a model Hamiltonian $\hat{H}$ composed of
the phenomenological Woods-Saxon potential and
the schematic separable-type interaction,
which has been also utilized in Refs.~\cite{TS12,TSD13}.
Its precise form is given in part~I
and we will not repeat it here.

When the projected wave function in Eq.~(\ref{eq:wfProj}) is obtained,
it is straightforward to calculate
the electromagnetic transition probabilities~\cite{RS80}.
No effective charge is used for the calculation of $B(E2)$
because the full model space is employed without any kind of core.
The effective spin $g$-factor of $0.7\times g_{s,{\rm free}}$ is adopted
for both neutrons and protons for the calculation of $B(M1)$.
In this way there is no ambiguity for the calculation of
these transition probabilities.

The product-type mean-field wave function
with the pairing correlations, $|\Phi \rangle$ in Eq.~(\ref{eq:wfProj}),
is generated by the mean-field Hamiltonian $\hat{h}_{\rm mf}$
composed of the deformed Woods-Saxon potential
and the monopole-type pairing potential,
where the pairing potential has the form factor of
the derivative of the Woods-Saxon potential, see part~I for details.
The deformation in the body-fixed frame
is specified with respect to the equi-potential surface
at the half depth for the Woods-Saxon potential
with the usual radius parameterization,
\begin{equation}
 R(\theta,\varphi)=
 R_0 \,c_v(\{\alpha\})
 \bigg[ 1+\sum_{\lambda\mu}\alpha^*_{\lambda\mu}
 Y_{\lambda\mu}^{}(\theta,\varphi) \bigg],
\label{eq:surf}
\end{equation}
with the quantity $c_v(\{\alpha\})$ that
guarantees the volume-conservation condition.
In the present work, we employ $\lambda=2$ and 4 deformations with
the parameters $(\beta_2,\beta_4,\gamma)$,
where the so-called Lund convention~\cite{BR85} is used
for the sign of triaxiality parameter $\gamma$, and therefore,
for example,
$\langle x^2 \rangle < \langle y^2 \rangle < \langle z^2 \rangle$
for $0^\circ < \gamma < 60^\circ$.
Here $\langle x^2 \rangle$ etc. are abbreviated notations of
${\displaystyle \Bigl\langle \sum_{i=a}^A \bigl(x^2\bigr)_a\Bigr\rangle}$ etc.,
which will be also used in the following discussions.

It is worthwhile mentioning that
the triaxiality parameter in the Woods-Saxon potential,
$\gamma\equiv\gamma({\rm WS})$,
and the corresponding parameter in the Nilsson potential,
$\gamma({\rm Nils})$, are somewhat different from
that of the density distribution for the mean-field state,
$\gamma({\rm den})$, which is defined by
\begin{equation}
 \gamma({\rm den})\equiv
 \tan^{-1}\biggl[-\frac{\sqrt{2}\langle Q_{22}\rangle}{\langle Q_{20}\rangle}
  \biggr] ,
\label{eq:gammaden}
\end{equation}
where $Q_{2\mu}$ is the quadrupole operator;
see Ref.~\cite{SSM08} for the precise definitions of the various
$\gamma$ parameters and discussion related to them.
Although the difference between these quantities,
e.g., $\gamma({\rm WS})$ and $\gamma({\rm den})$,
are not so large as in the case of the wobbling motion
for the triaxial superdeformed nuclei,
they are still sizable and one has to be careful
for discussing the triaxial deformation.

One of the interesting quantities studied in part~I
and also in Ref.~\cite{TS16} is
the expectation value of the angular-momentum vector
in the body-fixed frame specified by the mean-field,
from which the projection is performed.
Following the previous work~\cite{TS16}, we define
the expectation value of each component of
the angular-momentum vector in the intrinsic frame
for the projected eigenstate $\alpha$ in the following way,
\begin{equation}
 (\!( J^2_i )\!)_\alpha
 \equiv \sum_{KK'} f^{I*}_{K,\alpha}\,
 \langle IK|J^2_i|IK'\rangle\,f^I_{K',\alpha},
\label{eq:exJJ}
\end{equation}
where the index $i=x,y,z$ denotes the axis
specified by the deformed mean-field wave function $|\Phi\rangle$,
and the $(f^I_{K,\alpha})$ are
the properly orthonormalized amplitudes~\cite{RS80},
which are defined with the help of the square-root matrix of the norm kernel by
\begin{equation}
f^I_{K,\alpha}=\sum_{K'}
\bigl(\sqrt{{\cal N}^I}\,\bigr)_{K,K'}\, g^I_{K',\alpha}.
\label{eq:normfNocm}
\end{equation}
Needless to say, the purely algebraic quantity $\langle IK|J^2_i|IK'\rangle$,
e.g., $\langle IK|J^2_z|IK'\rangle=\delta_{KK'}K^2$,
should be calculated in the intrinsic frame with $[J_x,J_y]=-i\hbar J_z$ etc.
The microscopic geometrical information is contained
in the amplitude $f^I_{K,\alpha}$.
A more microscopic definition by using the mean-field wave function
is necessary to obtain the neutron and proton contributions separately;
they are evaluated by ($\tau={\rm n,p}$)
\begin{equation}
 \langle\!\langle {J_i^{(\tau)2}} \rangle\!\rangle_\alpha
 \equiv {\rm Re}\biggl[\sum_{KK'} g^{I*}_{K,\alpha}\,
 \langle \Phi|{J_i^{(\tau)2}} \hat{P}_{KK'}^I|\Phi\rangle\,
 g^I_{K',\alpha}\biggr],
\label{eq:exJJm}
\end{equation}
which are shown to be consistent with the definition of
the total expectation value in Eq.~(\ref{eq:exJJ}); i.e.,
$\langle\!\langle {J_i^{({\rm n})2}} \rangle\!\rangle_\alpha
+\langle\!\langle {J_i^{({\rm p})2}} \rangle\!\rangle_\alpha
\approx (\!({J_i^2})\!)_\alpha$ in a very good approximation,
see the discussion in Appendix of Ref.~\cite{TS16}.

\section{Application to a chiral doublet band}
\label{sec:chiral}

The possible existence of chiral doublet bands was first pointed out
by Frauendorf and Meng in Ref.~\cite{FM97},
and has been explored experimentally since then;
see, e.g., Refs.~\cite{SKC01,KSC03}.
This interesting rotational motion is characteristic
for triaxially deformed nuclei.
As it was already discussed in part~I of the study of
nuclear wobbling motion,
there are three distinct directions in the body-fixed frame
of triaxially deformed nucleus.
In the present work, we choose the intrinsic coordinate system that satisfies
$\langle x^2 \rangle < \langle y^2 \rangle < \langle z^2 \rangle$;
namely, $0 < \gamma < 60^\circ$,
and the short, medium, and long axes are $x$, $y$, and $z$ axes, respectively.
If there are three different kinds of angular-momentum vectors,
which favor aligning their vectors along these three principal axes,
the three vectors are aplanar in the intrinsic frame.
In such a situation, the symmetry of the ``handedness'' is broken;
i.e., whether these three angular-momentum vectors are
right-handed or left-handed in the $xyz$ intrinsic coordinate
is chosen by the selfconsistent mean-field.
Just as in the case of the parity doublet,
two almost degenerate ${\mit\Delta I}=1$ rotational bands
are expected as different linear combinations of
the right- and left-handed states, which appear as the chiral doublet bands,
see Sec.~\ref{sec:selrule} for details.

A prototype example,
which was considered in Ref.~\cite{FM97} and in Ref.~\cite{KSH04}
is the odd-odd nucleus
with an odd proton sitting in the high-$j$ particle-like orbit
and an odd neutron in the high-$j$ hole-like orbit (or vice versa).
The high-$j$ particle-like orbit tends to align its angular-momentum vector
along the short ($x$) axis,
while the high-$j$ hole-like orbit tends to align along the long ($z$) axis.
Moreover, the collective angular-momentum prefers to align to
the axis with the largest moment of inertia,
which is the medium ($y$) axis for the irrotational-like moments of inertia.
Thus an aplanar angular-momentum geometry, i.e., chiral geometry,
is expected to appear at the critical spin $I_{\rm c}$
Below $I_{\rm c}$ the collective angular-momentum lies in the $xz$ plane,
see discussion of transverse wobbling~\cite{Fra17}.
In the present work we study the nucleus $^{128}$Cs,
in which the odd proton (neutron) occupies the quasiparticle state
whose main component is particle-like (hole-like) $h_{11/2}$ orbit,
and discuss how the chiral geometry comes about.
Especially, it is shown that the ideal situation considered in Ref.~\cite{KSH04}
is indeed realized in our microscopic calculations.
In order to demonstrate that the appearance of such an ideal chiral doublet band
is not very rare in the calculation,
we also briefly discuss another example, $^{104}$Rh,
in which the odd proton (neutron) occupies the quasiparticle state
whose main component is hole-like $g_{9/2}$ (particle-like $h_{11/2}$) orbit.

In the following we investigate the chiral doublet bands
by the fully microscopic framework of the angular-momentum-projection approach
in contrast to the original work~\cite{FM97,KSH04},
where the macroscopic model of the triaxial-rotor
coupled to a particle and a hole is employed.
The calculational procedure is the same as in part~I.
The calculations are performed within the isotropic harmonic oscillator basis
and the basis states are truncated up to the maximum oscillator shells,
$N_{\rm osc}^{\rm max}=12$.
As it was explained in detail in part~I
the monopole-type pairing force strengths are determined to reproduce
the even-odd mass differences for the neighboring even-even nuclei,
and the average of those is adopted for the odd-odd nucleus.
In the present calculation the average pairing gaps
for both neutrons and protons are
calculated selfconsistently using the strengths thus determined.
Since we do not intend to reproduce the experimental data
but perform explorational calculations,
we arbitrarily choose an appropriate value for
the deformation parameter $\beta_2$, and $\beta_4=0.0$ for simplicity,
to obtain an ideal chiral geometry.
As for the triaxial deformation
$\gamma({\rm WS})=30^\circ$ is adopted for the Woods-Saxon mean-field.

\subsection{Chiral geometry and selection rules for transition rates}
\label{sec:selrule}

Before showing the result of our angular-momentum projection calculation,
we briefly discuss how the chiral geometry is realized and
what is expected for it according to Refs.~\cite{FM97,KSH04};
see also the review articles~\cite{Fra01,SK17,Fra17}.
In the simple classical model,
where a particle and a hole angular momenta,
$j_{\rm p}$ and $j_{\rm h}$, align along the short ($x$) axis
and the long ($z$) axis , respectively,
the trajectory of the angular-momentum vector
($J_x,J_y,J_z$) is given by the intersection
of the sphere and the shifted ellipsoid, which are described by
the equations,
\begin{equation}
\left\{\begin{array}{l}
 J_x^2+J_y^2+J_z^2=I(I+1), \vspace*{2mm}\cr
 {\displaystyle
 \frac{(J_x-j_{\rm p})^2}{2{\cal J}_x} +\frac{J_y^2}{2{\cal J}_y}
 +\frac{(J_z-j_{\rm h})^2}{2{\cal J}_z}=E }, \end{array}\right.
\label{eq:paroteq}
\end{equation}
representing the angular-momentum conservation, $I$,
and the rotor model energy, $E$, respectively.
The quantities, ${\cal J}_x$, ${\cal J}_y$ and ${\cal J}_z$,
are the moments of inertia of the core nucleus in the body-fixed frame,
and it is assumed that the medium axis has the largest inertia,
i.e., ${\cal J}_y> {\cal J}_x,\,{\cal J}_z$.
At low spins the trajectory of the lowest energy state is mainly confined
in the $xz$ principal plane, $J_x\approx j_{\rm p}$, $J_z\approx j_{\rm h}$,
and $J_y \approx 0$, and in the first excited state the angular-momentum
vector vibrates with respect to this plane,
i.e., the so-called the chiral vibration~\cite{SKC01}.
This chiral vibrational excitation has been studied microscopically
by the quasiparticle random-phase approximation in Ref.~\cite{ADF11}.

When the spin increases and exceeds the critical spin~\cite{ODD04},
\begin{equation}
I_{\rm c}= \left[\left(
 \frac{j_{\rm p}{\cal J}_y}{{\cal J}_y-{\cal J}_x}\right)^2
 +\left(\frac{j_{\rm h}{\cal J}_y}{{\cal J}_y-{\cal J}_z}\right)^2\right]^{1/2},
\label{eq:critI}
\end{equation}
the chiral symmetry is broken in the yrast states;
i.e., the aplanar angular-momentum geometry is realized
giving the lowest two degenerate solutions,
the right-handed (e.g., $J_y >0$), and
the left-handed (e.g., $J_y <0$) ones, which we denote by
$|r \rangle$ and $|l \rangle$, respectively.
They are related by $|l \rangle = {\cal T}\hat{R}_y(\pi)|r \rangle$,
where the operation  ${\cal T}$ is the time-reversal transformation
and $\hat{R}_y(\pi)$ is the $\pi$-rotation about the $y$ axis.
The mean-field solution, which shows the aplanar chiral geometry,
was obtained for the first time in Ref.~\cite{VFD00}
by the microscopic framework of
shell-correction tilted axis cranking approach.
With the same approach, the transition to the chiral geometry and
its critical spin value were investigated
from the microscopic view point in Ref.~\cite{ZGN03}
in comparison with the experimental data.

There is tunneling effect between the two solutions,
$|r \rangle$ and $|l \rangle$,
and the quantum mechanical eigenstates are obtained by
the linear combinations,
\begin{equation}
 |+\rangle=\frac{1}{\sqrt{2}}\bigl(|r \rangle+|l \rangle\bigr),\qquad
 |-\rangle=\frac{i}{\sqrt{2}}\bigl(|r \rangle-|l \rangle\bigr),
\label{eq:chrleigs}
\end{equation}
which are interpreted as the chiral doublet states
just like the parity doublet states.
Note that the partner states in Eq.~(\ref{eq:chrleigs}) are constructed
for each spin value, $\cdots$, $I-1$, $I$, $I+1$, $\cdots$.
Now let us consider the electromagnetic transition rates such as $E2$ and $M1$.
Since the photon with these low multipolarities cannot turn
the angular-momentum vector from the right- to the left-hand position,
the overlap $\langle l|\widehat{\cal M}|r \rangle$ of the transition operator
$\widehat{\cal M}$ is essentially vanishing
after the static chiral geometry is realized.
Thus these transition rates satisfy the selection rules,
\begin{equation}
 B({\cal M};\,+\rightarrow +)\approx B({\cal M};\,-\rightarrow -),\qquad
 B({\cal M};\,-\rightarrow +)\approx B({\cal M};\,+\rightarrow -).
\label{eq:Btrchrl}
\end{equation}
Namely, both the in-band transitions and the out-of-band
for the pair of the doublet bands are the same.

\begin{figure}[!htb]
\begin{center}
\includegraphics[width=150mm]{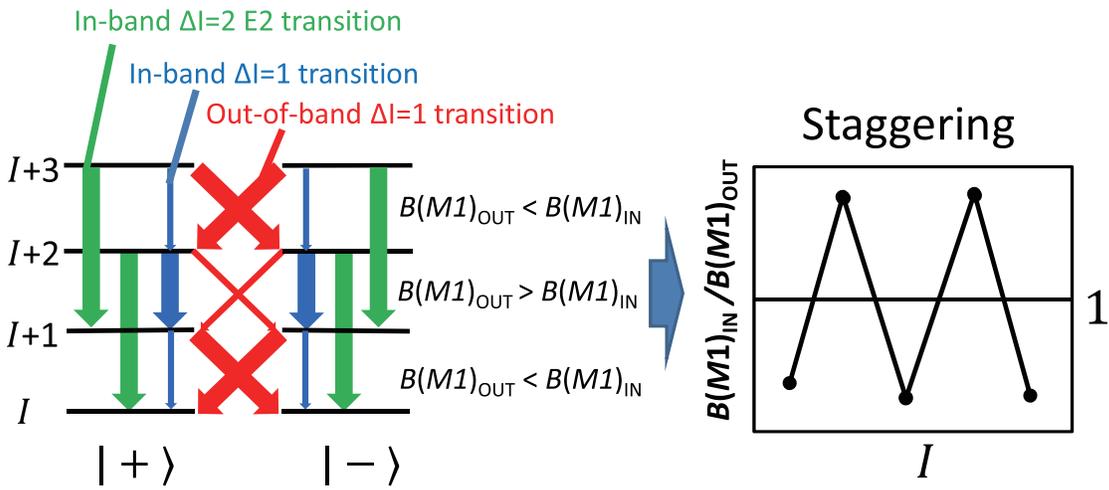}
\vspace*{-4mm}
\caption{(Color online)
Schematic figure representing the selection rules of the
electromagnetic transitions for the ideal chiral geometry
considered in Ref.~\cite{KSH04},
where the thick (thin) arrow denotes the large (small) transition rate.
}
\label{fig:selchrl}
\end{center}
\end{figure}

In Ref.~\cite{KSH04} an interesting typical case is considered,
which shows especially characteristic properties
for the $E2$ and $M1$ transition rates, when the chiral symmetry is broken.
Namely, the system is invariant with respect to the combined operation
of the $\pi/2$-rotation about the medium ($y$) axis
and an exchange of the valence neutron and proton;
this operation is called $\hat{A}$ hereafter,
and the eigenstates are classified by the eigenvalues $\pm 1$ of $\hat{A}$.
Within the simple model in Eq.~(\ref{eq:paroteq})
the system is $\hat{A}$-invariant if the moments of inertia
satisfy the condition, ${\cal J}_x={\cal J}_z$ and
the valence neutron and proton sit in the same high-$j$ orbit,
because the $\pi/2$-rotation about the $y$ axis
interchanges the $x$ and $z$ axes and at the same time
the exchange of valence neutron and proton
interchanges the particle and hole alignments $j_{\rm p}$ and $j_{\rm h}$.
Considering that the contribution of the valence neutron and proton
is almost negligible for the $E2$ operator
and that the $M1$ operator has approximate isovector character,
it has been shown that these transitions
are almost prohibited between the states with same eigenvalue of $\hat{A}$.
Moreover, chirality requires that
the partner states in Eq.~(\ref{eq:chrleigs}) at a given spin
have different eigenvalues of $\hat{A}$,
because the exchange between the valence neutron and proton
while keeping the direction of the rotor angular-momentum
changes the right- into left-handed states.
Taking into account the considerations on top of Eq.~(\ref{eq:Btrchrl}),
the selection rules for the $E2$ and $M1$ transition rates
inside and/or between the chiral doublet bands can be summarized
in Fig.~\ref{fig:selchrl};
see Ref.~\cite{KSH04} for more detailed discussions.
An especially interesting property is seen
in the ${\mit\Delta}I=1$ $E2$ and $M1$ transitions;
the large in-band and small out-of-band transitions and
the small in-band and large out-of-band transitions alternate with spin,
which can be more clearly observed
for the ratio of in-band and out-of-band transitions,
e.g., $B(M1)_{\rm in}/B(M1)_{\rm out}$,
as it is depicted schematically in the right part of Fig.~\ref{fig:selchrl}.

\begin{figure}[!htb]
\begin{center}
\includegraphics[width=120mm]{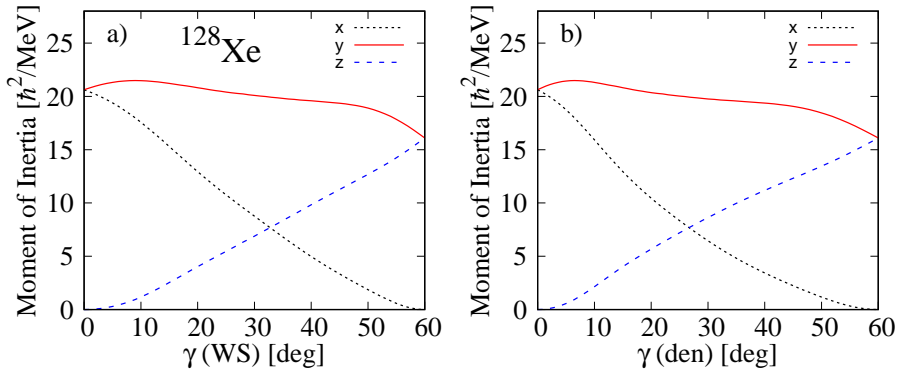}
\vspace*{-4mm}
\caption{(Color online)
Cranking moments of inertia of the three intrinsic axes,
$x$, $y$, and $z$, which are the short, medium, and long axes,
(denoted by dotted, solid, and dashed lines, respectively)
as functions of the triaxiality parameter $\gamma$
for the even-even core nucleus $^{128}$Xe of $^{128}$Cs.
The deformation parameters are $\beta_2=0.30,\beta_4=0.0$ and
the pairing gaps are $\Delta_{\rm n}=0.85$ MeV and $\Delta_{\rm p}=1.07$ MeV,
which are are employed for the study of chirality in $^{128}$Cs.
The $\gamma$ parameter of the Woods-Saxon potential is utilized in a) and
that of the density distribution, Eq.~(\ref{eq:gammaden}), in b).
}
\label{fig:momiXe}
\end{center}
\end{figure}

\begin{figure}[!htb]
\begin{center}
\includegraphics[width=120mm]{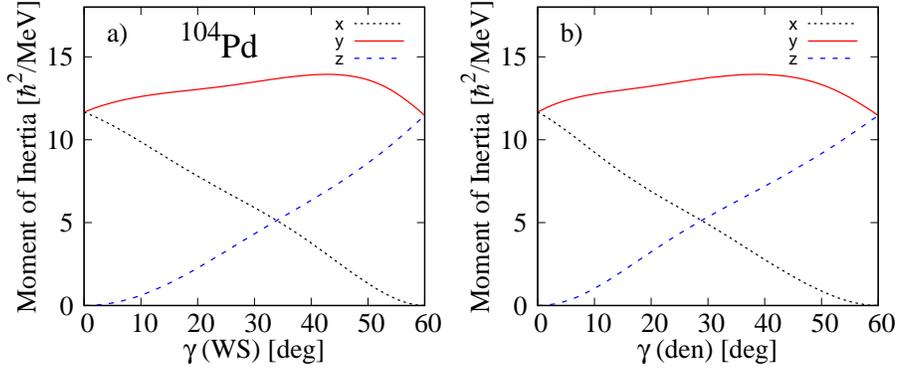}
\vspace*{-4mm}
\caption{(Color online)
The same as Fig.~\ref{fig:momiXe} but for the even-even core
nucleus $^{104}$Pd of $^{104}$Rh.
The deformation parameters are $\beta_2=0.25,\beta_4=0.0$ and
the pairing gaps are $\Delta_{\rm n}=0.95$ MeV and $\Delta_{\rm p}=0.76$ MeV,
which are employed for the study of chirality in $^{104}$Rh.
}
\label{fig:momiPd}
\end{center}
\end{figure}

In the following we will show that the ideal chiral geometry considered
in Ref.~\cite{KSH04} is indeed realized in our angular-momentum projection
calculation.  This is non-trivial because we do not introduce any kind of
rotor and/or valence nucleons explicitly in our fully microscopic framework.
However, it is instructive to see moments of inertia of
the even-even ``core''nuclei;
i.e., $^{128}$Xe for the odd-odd nucleus $^{128}$Cs
with an odd proton particle and an odd neutron hole,
and $^{104}$Pd for the odd-odd nucleus $^{104}$Rh
with an odd neutron particle and an odd proton hole.
Although we do not explicitly use the three moments of inertia
of the principal axes in our framework, their values are of interest.
They can be estimated by the cranking procedure; ${\cal J}_i
={\displaystyle \mathop{\rm lim}_{\omega_i \rightarrow 0}}
\langle J_i \rangle/\omega_i$, where $\omega_i$ is the cranking frequency
about the $i$-th axis ($i=x,y,z$) of the intrinsic frame.
Figures~\ref{fig:momiXe} and~\ref{fig:momiPd} display
the calculated cranking moments of inertia for $^{128}$Xe and $^{104}$Pd,
respectively,
as functions of the triaxiality parameter, $\gamma{\rm (WS)}$,
for the Woods-Saxon potential.
As it has been discussed for the wobbling motion
in part~I, the different definitions of the triaxiality parameter
give considerably different values~\cite{SSM08}.
Therefore, we show the same quantities
as functions of the triaxiality parameter, $\gamma{\rm (den)}$,
for the density distribution defined in Eq.~(\ref{eq:gammaden}),
although the differences are not so large as in the case of
the triaxial superdeformed states in part~I.
It should be mentioned that
the used mean-field parameters other than the triaxiality are those employed
for the analysis of the chiral doublet bands for $^{128}$Cs and $^{104}$Rh.
Therefore the calculated moments of inertia may not be very realistic
for $^{128}$Xe and $^{104}$Pd nuclei themselves.
The dependence of three moments of inertia on $\gamma{\rm (den)}$
resembles that of irrotational flow; see, e.g., Fig.~1 of part~I.
However, the relative values are considerably different;
at $\gamma=30^\circ$ ${\cal J}_y$ is larger than ${\cal J}_x={\cal J}_z$
by a factor 4 for the irrotational flow,
while the factor is 2.4--2.6 for the microscopic ones
in Figs.~\ref{fig:momiXe} and~\ref{fig:momiPd}.
This result is similar to the triaxial superdeformed state
in $^{163}$Lu studied in part~I.
We adopt the value $\gamma{\rm (WS)}=30^\circ$ for the analysis,
and it can be seen that the necessary condition,
${\cal J}_y> {\cal J}_x\approx{\cal J}_z$, is approximately satisfied
in both $^{128}$Cs and $^{104}$Rh.
With these cranking moments of inertia at $\gamma{\rm (WS)}=30^\circ$
and assuming full alignments for the $h_{11/2}$ and $g_{9/2}$ orbits,
the critical angular-momentum in Eq.~(\ref{eq:critI}) can be estimated
to $I_{\rm c}\approx 12.9$ for $^{128}$Cs
and $I_{\rm c}\approx 11.8$ for $^{104}$Rh.

\subsection{Chiral doublet band in $^{128}$Cs}
\label{sec:chiralCs}

In the course of our investigation we have found that
it is difficult to obtain the doublet bands
if the mean-field is cranked with finite rotational frequencies.
Therefore, we do not crank the mean-field
or just try to make infinitesimal cranking~\cite{TS16} with 10~keV frequencies
about three principal axes as it was studied in part~I.
If the mean-field is constructed without cranking there is an ambiguity
related to the fact that the single-particle states are doubly degenerate
(i.e. the Kramers degeneracy),
which was already discussed in part~I for the odd nucleus $^{163}$Lu.
These doubly degenerate states are usually classified by the signature,
which is the symmetry with respect to the $\pi$ rotation
about one of the intrinsic coordinate axes.
We choose the $x$ axis and classify the single-particle orbits
by $\hat{R}_x(\pi)$; namely a {\it favored} signature state $\alpha$
and its conjugate {\it unfavored} state $\bar{\alpha}=-\alpha$.
In the case of odd nucleus there is no ambiguity,
because the mean-field state with an odd particle in the $\bar{\alpha}$ state
is obtained from
the mean-field state with an odd particle in the $\alpha$ state
by the rotation $\hat{R}_x(\pi)$, and therefore the result of
the angular-momentum projection from these two states is exactly the same.
For an odd-odd nucleus, however, there are four possible configurations
for occupying the odd neutron and odd proton; i.e.,
\begin{equation}
\mbox{(a) }(\alpha_\nu,\alpha_\pi),\quad
\mbox{(b) }(\bar{\alpha}_\nu,\alpha_\pi),\quad
\mbox{(c) }(\alpha_\nu,\bar{\alpha}_\pi),\quad
\mbox{(d) }(\bar{\alpha}_\nu,\bar{\alpha}_\pi) \,.
\label{eq:sig4class}
\end{equation}
Among them the configuration (d) is obtained from (a) by $\hat{R}_x(\pi)$,
and (c) from (b) by $\hat{R}_x(\pi)$, but the configurations
(a) and (b) are independent for the angular-momentum projection calculation.
We have numerically confirmed this fact; i.e., the result of projection
from the blocked configuration (d) is exactly the same as that from (a),
and the result from (c) is the same as that from (b).
However, the results of projection from (a) and from (b) are different,
although the differences are found to be very small.
One possible way to get rid of the ambiguity is to mix
these two independent configurations (a) and (b) in Eq.~(\ref{eq:sig4class}).
We will discuss this point in the following.
Practically we use extremely small cranking frequency
$\omega_x=10^{-10}$ MeV/$\hbar$ and block the lowest
quasiparticle state to generate the configuration $\alpha$ and
block the second lowest state for the configuration $\bar{\alpha}$.
In the following study of $^{128}$Cs the blocking of
the negative-parity quasiparticle-state originating from
the $h_{11/2}$ orbit has been done for both neutron and proton.

\begin{figure}[!htb]
\begin{center}
\includegraphics[width=75mm]{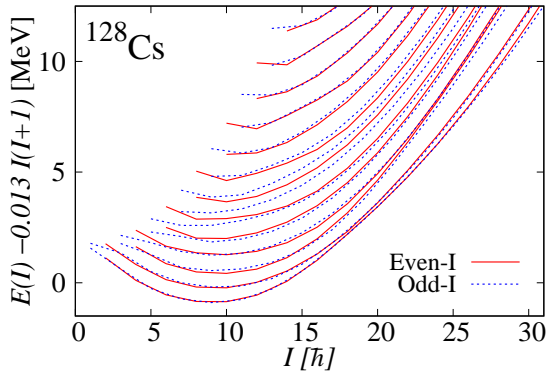}
\vspace*{-4mm}
\caption{(Color online)
Energy spectrum for $^{128}$Cs calculated by
the angular-momentum-projection method from the non-cranked mean-field
with the configuration (a) in Eq.~(\ref{eq:sig4class}).
A rigid-rotor reference energy $0.013\,I(I+1)$ MeV is subtracted.
}
\label{fig:sCsnocr}
\end{center}
\end{figure}

The ideal chiral geometry for doublet band
does not always appear in the calculation;
we need to choose proper deformation parameter.
We found that it appears at the deformation parameter $\beta_2=0.30$
without hexadecapole deformation; thus we have chosen
$\beta_2=0.30$, $\beta_4=0.0$ and $\gamma=30^\circ$
for the Woods-Saxon mean-field
in the following investigation of the chiral doublet band in $^{128}$Cs.
Note that $\gamma=\gamma({\rm WS})=30^\circ$ corresponds to
$\gamma({\rm den})=24.0^\circ$ in this case.
Then, the average pairing gaps calculated selfconsistently
are $\Delta_{\rm n}=0.85$ MeV
and $\Delta_{\rm p}=1.07$ MeV for neutrons and protons, respectively.
It should be mentioned that the adopted value, $\beta_2=0.30$,
is considerably larger than the ordinary used value,
$\beta_2 \approx 0.15-0.20$,
in the nuclear region around $^{128}$Cs.
The resultant rotational spectrum is displayed in Fig.~\ref{fig:sCsnocr},
where the angular-momentum projection is performed
from the non-cranked mean-field
with the configuration (a) in Eq.~(\ref{eq:sig4class}).
We have done the same calculation with the configuration (b),
but the result is very similar and is not shown.
In this and the following figures the even-$I$ and odd-$I$ sequences
of the band are connected by the solid and dashed lines, respectively,
and a rigid-rotor reference energy $0.013\,I(I+1)$ MeV is
subtracted to see more clearly the degeneracy of the bands.
This reference energy is selected such that the experimentally
observed yrast band is almost flat; see Fig.~\ref{fig:sCsmixed} below.
At first sight the excitation spectrum of the multiple-band structure
is similar to the wobbling bands in $^{163}$Lu studied in part~I.
However, the yrast wobbling band in $^{163}$Lu has
the spin values $I-1/2$ being even, and the first excited band has
the spin values $I-1/2$ being odd, etc.; i.e.,
the bands are composed of the ${\mit\Delta}I=2$ states and
the signature of the multiple wobbling band
alternates with increasing energy.
In the present case of the multiple bands in $^{128}$Cs,
the even-$I$ and odd-$I$ bands are almost degenerate and
form ${\mit\Delta}I=1$ rotational bands.
This is because the odd proton aligns its angular-momentum vector
along the short ($x$) axis,
while the odd neutron-hole aligns its angular-momentum vector
along the long ($z$) axis,
and consequently the signature symmetry is strongly broken
already at low spins.  This is confirmed later by looking at
the expectation values of the intrinsic angular-momentum vector
in Fig.~\ref{fig:jCsmixed} below.
What is important for the spectrum of $^{128}$Cs in Fig.~\ref{fig:sCsnocr}
is that the energies of the lowest (yrast) and the second lowest (yrare)
${\mit\Delta}I=1$ bands become very close
in the spin range, $15 \ltsim I \ltsim 25$,
which can be well interpreted as the chiral doublet band~\cite{FM97}.
The calculated minimum energy-difference
between the two bands is about 120 keV at $I=18$.
The estimated critical spin $I_{\rm c}\approx 13$ in Sec.~\ref{sec:selrule}
is slightly smaller than the spin
where the two bands become almost degenerate.

\begin{figure*}[!htb]
\begin{center}
\includegraphics[width=155mm]{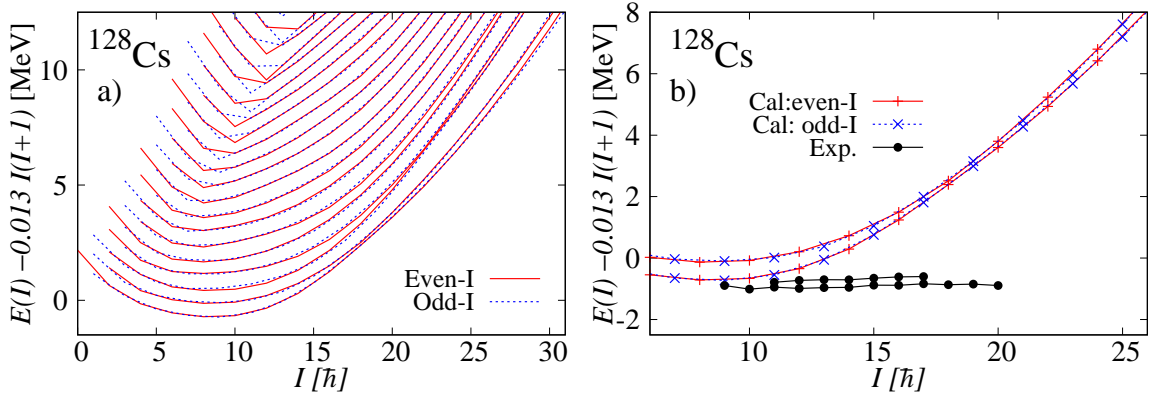}
\vspace*{-4mm}
\caption{(Color online)
Left panel: Energy spectrum for $^{128}$Cs calculated by
the angular-momentum-projected configuration-mixing method
with the two mean-field configurations (a) and (b) in Eq.~(\ref{eq:sig4class}).
Right panel: Comparison of the calculated and experimental
chiral doublet bands in $^{128}$Cs.
Experimental data are taken from Ref.~\cite{KSC03}.
}
\label{fig:sCsmixed}
\end{center}
\end{figure*}

There are two possible configurations for
the odd-odd nucleus, those of (a) and (b) in Eq.~(\ref{eq:sig4class}).
Although the resultant spectra obtained from the two configurations
are rather similar,
we have performed the projected configuration-mixing by
including these two configurations in order to obtain the unambiguous result,
which is shown in the left panel of Fig.~\ref{fig:sCsmixed}.
Comparing with the result of Fig.~\ref{fig:sCsnocr},
the higher-lying spectrum changes considerably;
especially the excitation energies become lower and the level density
of the excited bands is higher than those without configuration-mixing.
However, it should be stressed that the yrast and the yrare bands,
which are almost degenerate and so interpreted as chiral doublet bands,
are almost the same as without configuration-mixing
in Fig.~\ref{fig:sCsnocr}.
However, the agreement with the experimental data is
not very good as seen in the right panel of Fig.~\ref{fig:sCsmixed}.
The calculated bands become degenerate at about $I\approx 15$,
while experimentally two bands come together already at $I\approx 11$.
Moreover, the slope of the bands is too steep in the calculated bands;
namely the moments of inertia are too small compared
with the experimental data,
which is similar to the calculation of the wobbling band
in $^{163}$Lu studied in part~I.

As it has been pointed out in Ref.~\cite{TS16} infinitesimal cranking
quite often improves the calculated moments of inertia.
Therefore, we have tried to apply it also to the present case of $^{128}$Cs
with frequencies $\omega_x=\omega_y=\omega_z=0.01$ MeV$/\hbar$,
where the projection is performed from the single configuration of
the infinitesimally cranked mean-field state that is
composed of the lowest energy quasiparticle state
for both the odd neutron and proton being blocked.
We have found that the result of calculation is very similar to that of
mixing the two configurations in Fig.~\ref{fig:sCsmixed},
and it is not shown.
Namely, the moments of inertia of calculated rotational bands
are not improved in this particular case.
This result indicates that the time-odd components of the wave function
induced by the infinitesimal cranking do not contribute to increase
the moments of inertia in this case,
although they contain the two different configurations
(a) and (b) in Eq.~(\ref{eq:sig4class}) and their mixing effect.

\begin{figure*}[!htb]
\begin{center}
\includegraphics[width=155mm]{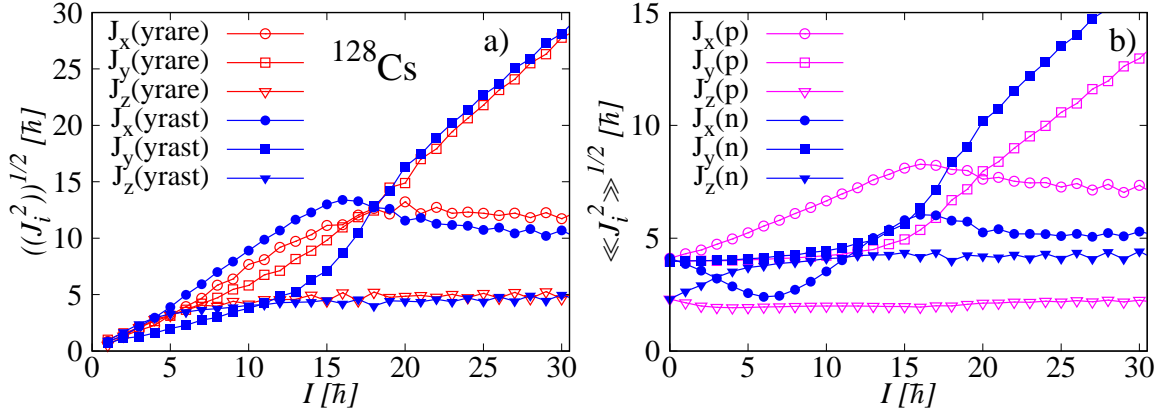}
\vspace*{-4mm}
\caption{(Color online)
The calculated expectation values of the angular-momentum vector
in the intrinsic frame for the configuration-mixed calculation
of $^{128}$Cs corresponding to the spectrum in Fig.~\ref{fig:sCsmixed}.
The left panel shows the expectation values of the total vector
for the yrast (filled symbols) and yrare (open symbols)
${\mit\Delta}I=1$ bands, while the right panel shows
the neutron (filled symbols) and proton (open symbols) contributions
in Eq.~(\ref{eq:exJJm}) separately for the yrast band.
Note that the $x$, $y$, and $z$ axes are the short, medium, and long axes,
respectively.
}
\label{fig:jCsmixed}
\end{center}
\end{figure*}

In order to study the dynamics of the angular-momentum vector,
the expectation values of its components in the intrinsic frame,
calculated by Eq.~(\ref{eq:exJJ}) for the yrast and yrare bands,
are shown in the left panel of Fig.~\ref{fig:jCsmixed}.
Here the results of the configuration-mixed calculation corresponding
to Fig.~\ref{fig:sCsmixed} are displayed but the results are
qualitatively similar for other cases.
As it is shown in the left panel, all the three components of
the expectation values of the intrinsic angular-momentum vector
are non-negligible.
In the lower spin region, $I \ltsim 8$,
the dominant components are those along the short ($x$) and the long ($z$) axes
for the yrast band.
As the spin increases, the component of the medium ($y$) axis,
which is the axis with the largest moment of inertia
of the core nucleus (see Fig.~\ref{fig:momiXe}) quickly increases.
In the spin range $15 \ltsim I \ltsim 25$
the yrast and yrare have nearly the same geometries,
which is characteristic for the chiral regime.
The largest component of the angular-momentum vector changes
from being along the $x$ axis to along the $y$ axis at $I\approx 18$,
which roughly corresponds to the critical angular-momentum of
the appearance of the chiral doublet band in Fig.~\ref{fig:sCsmixed}.
This correspondence between the critical angular-momentum
and the transition of direction of the angular-momentum vector
in the intrinsic frame has also been discussed for the wobbling motion
in the $^{163}$Lu nucleus studied in part~I.

\begin{figure*}[!htb]
\begin{center}
\includegraphics[width=120mm]{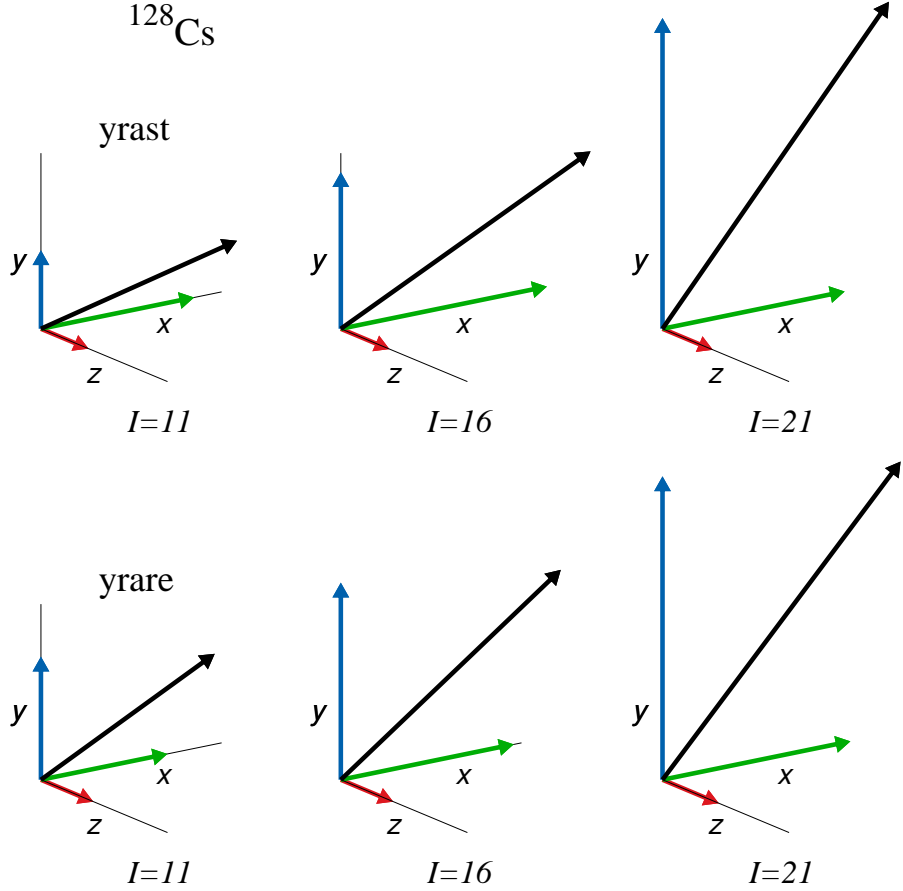}
\vspace*{-4mm}
\caption{(Color online)
Angular-momentum vectors in the intrinsic frame
for the $I=11$, 16, and 21 yrast (upper panel) and yrare (lower panel)
states in $^{128}$Cs
according to the expectation values shown in Fig.~\ref{fig:jCsmixed}.
}
\label{fig:arwsCs}
\end{center}
\end{figure*}

It can be also seen in Fig.~\ref{fig:jCsmixed}
that the $y$ component of the yrast band is considerably
smaller than that of the yrare band for $I \ltsim 17$,
which suggests that the vector of the yrast band stays
near the $xz$ principal plane,
while the vector of the yrare band goes back and forth
with respect to this plane;
i.e., the system is in the regime of chiral vibration~\cite{SKC01}.
Note that the quantities
$\langle\!\langle {J_i^2} \rangle\!\rangle$
in Eq.~(\ref{eq:exJJ}) include such effect of angular-momentum fluctuations.
In contrast, the three components of angular-momentum vectors
for a pair of the yrast and yrare bands are similar at $I \gtsim 18$,
and form the aplanar configuration, i.e., the system is in the regime
of the static chirality; this situation is exactly what is expected for
the chiral doublet band to appear.
The transition from the regime of the chiral vibration
to that of the static chirality occurs gradually.
A similar transition expected for the transverse wobbling and
related to the direction of the angular-momentum vector
in the intrinsic frame has been discussed in part~I.
How the total angular-momentum vector changes the direction
is shown pictorially in Fig.~\ref{fig:arwsCs}
according to the calculation in Fig.~\ref{fig:jCsmixed}:
The directions of two vectors for the yrast and yrare bands
are rather different at $I=11$; in fact that of the yrare band vibrates
with respect to the $xz$ principal plane.
At $I=21$ the two vectors points almost to the same direction,
and the yrast and yrare bands can be interpreted as a pair of the doublet band.
The $x$ and $z$ components stays almost constants at $I \gtsim 18$ and
the $y$-component becomes dominant at higher spins, $I \gtsim 30$.

In the chiral regime, the particle-like proton quasiparticle
aligns its angular-momentum along the $x$ axis and
the hole-like neutron quasiparticle
aligns its angular-momentum along the $z$ axis,
while the collective angular-momentum vector is mainly along the $y$ axis.
To see how neutrons and protons contribute to the expectation values
of the angular-momentum vector,
the neutron and proton contributions for the yrast band
are depicted separately in the right panel of Fig.~\ref{fig:jCsmixed}.
The contribution of neutrons or of protons cannot be
calculated by Eq.~(\ref{eq:exJJ});
one needs to look into the microscopic wave function explicitly,
and Eq.~(\ref{eq:exJJm}) should be used instead.
It is seen in the right panel of Fig.~\ref{fig:jCsmixed} that indeed
the dominant contribution to the $x$ component comes from proton and
that to the $z$ component from neutron,
while both neutron and proton contribute to the $y$ component
as expected for collective angular-momentum.
Note that for increase of the $x$ component in the lower spin $I \ltsim 16$,
both neutron and proton contribute, which suggests that there is
non-negligible amount of collective angular-momentum for it.
This is consistent with the ``classical'' model in Eq.~(\ref{eq:paroteq}).
Namely, collective angular-momentum increases in the $xz$ plane
below the critical spin $I_{\rm c}$, while its $y$ component starts
to increase after $I_{\rm c}$ with constant $xz$ components.
Thus, the ideal situation of the chiral geometry is realized in this case
by our fully microscopic angular-momentum-projection calculation.
A similar analysis of the expectation values of angular-momentum vectors
has been performed within the particle-rotor model in Refs.~\cite{QZM09,QZW09},
which allows one to discriminate between collective core and quasiparticle
angular-momenta in contrast to our microscopic analysis.

\begin{figure*}[!htb]
\begin{center}
\includegraphics[width=155mm]{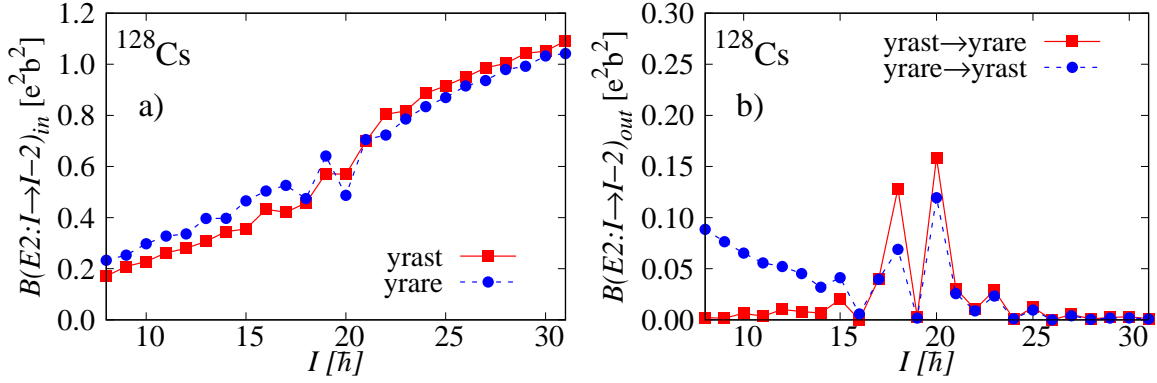}
\vspace*{-4mm}
\caption{(Color online)
The calculated $B(E2:I\rightarrow I-2)$ values for
the yrast band (solid lines) and the yrare band (dashed lines)
as functions of spin in $^{128}$Cs.
The left and right panels show the in-band and out-of-band values,
respectively.  Note that the ordinate scale in the right panel is different
from that in the left panel.
Shown are the results of the configuration-mixed projection calculation
corresponding to Fig.~\ref{fig:sCsmixed}.
}
\label{fig:tRCsmixed}
\end{center}
\end{figure*}

One of the merits of the angular-momentum-projection method is
that the electromagnetic transition rates between any eigenstates
can be calculated straightforwardly.
In the present work, we concentrate on the in-band and out-of-band
transitions for the two lowest ${\mit\Delta}I=1$ bands,
the yrast and the yrare bands, obtained by configuration-mixing
of the two configurations (a) and (b) in Eq.~(\ref{eq:sig4class}).
The $I \rightarrow I-2$ rotational $E2$ transition rates
are shown in Fig.~\ref{fig:tRCsmixed}.
The in-band transitions are always large.
In fact the large $B(E2:I\rightarrow I-2)$ values are used
to define each rotational band.
The in-band transition rates are similar for
the pair of the yrast and yrare bands,
where those of the yrare band are slightly larger at lower spins, $I \le 17$,
while those of the yrast band are slightly larger at higher spins, $I \ge 22$.
In contrast, the out-of-band transition probabilities are generally small,
although they are non-negligible in $17 \le I \le 23$,
where the splitting of the energies of the two bands are smallest and
the mixing between them is expected.
Note that the out-of-band transitions
from the yrare to the yrast at low spins, $I \le 15$, are not so small,
while those from the yrast to the yrare are very small,
which is characteristic for chiral vibrations.
The increasing behavior of the in-band $E2$ transitions results from
the change of direction of the angular-momentum vector in the intrinsic frame
for the fixed deformation of the mean-field.
In the semiclassical approximation the $B(E2)$ values
are proportional to $\bigl|\langle x_j^2-x_k^2 \rangle\bigr|^2$ for
the rotation about the $i$-th principal axis ($ijk$-cyclic), and
for rotation about the tilted axis the transition amplitudes
are given by the linear combination of these moments
$\langle x_j^2-x_k^2 \rangle$ depending on the angles
of the angular-momentum vector (c.f. the formulas in Refs.~\cite{Fra93,Fra00}).
In the present case, $0 < \gamma < 60^\circ$, the moment
$\langle z^2-x^2 \rangle$ is largest and the maximum value of
$B(E2)$ is expected for the rotation about the $y$ axis,
which is realized at much higher spins,
see the left panel of Fig.~\ref{fig:jCsmixed}.
This increase of the rotational $B(E2)$ values has been also seen
in the particle-rotor model calculation of Ref.~\cite{QZW09}.
However, the measured rotational $B(E2)$ values~\cite{GSP06} do not show
such increase with spin; they even slightly decrease
at highest spins observed.
Moreover, the calculated $B(E2)$ values are about factor two
larger than the measured values; this is because the value of $\beta_2$
employed in the present study is too large as mentioned previously.
Therefore, we do not attempt to make detailed comparison with experimental data
except for the $B(M1)_{\rm in}/B(M1)_{\rm out}$ ratio.

\begin{figure*}[!htb]
\begin{center}
\includegraphics[width=155mm]{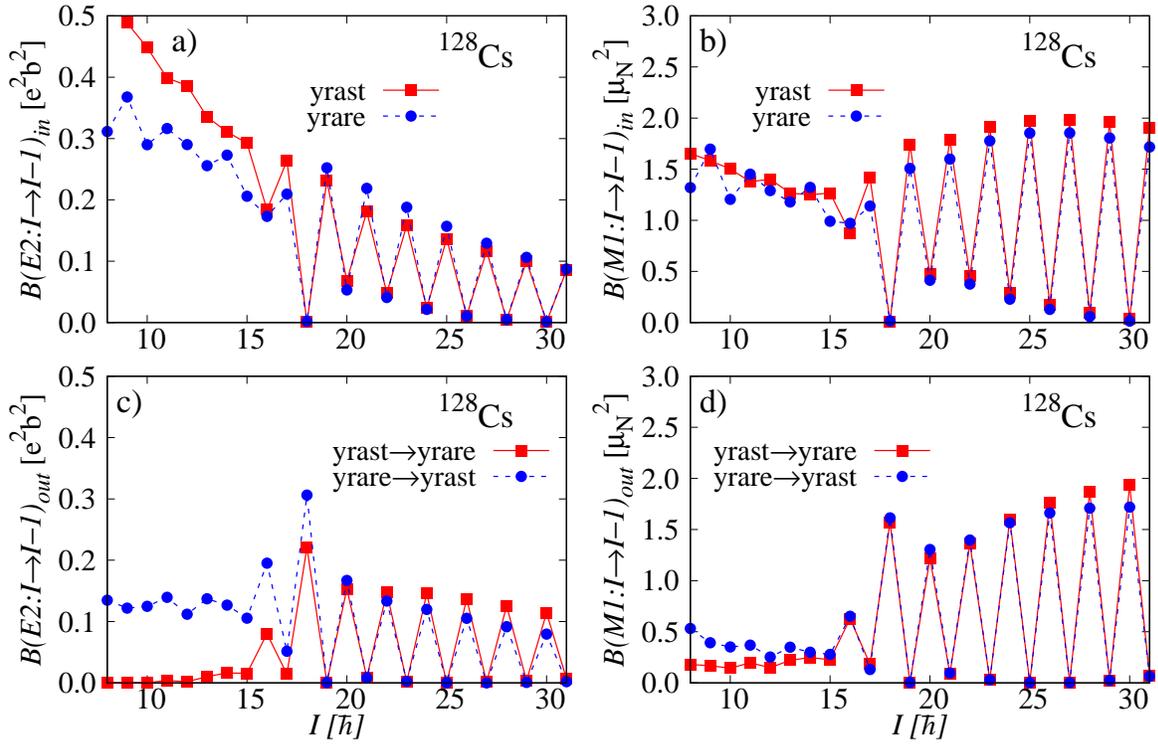}
\vspace*{-4mm}
\caption{(Color online)
The calculated
$B(E2:I\rightarrow I-1)$ and $B(M1:I\rightarrow I-1)$ values
for the yrast band (solid lines) and the yrare band (dashed lines)
as functions of spin in $^{128}$Cs.
The upper (lower) panels show the in-band (out-of-band) transition rates.
These are the results of the configuration-mixed projection calculation
corresponding to Fig.~\ref{fig:sCsmixed}.
}
\label{fig:tEMCsmixed}
\end{center}
\end{figure*}

\begin{figure*}[!htb]
\begin{center}
\includegraphics[width=155mm]{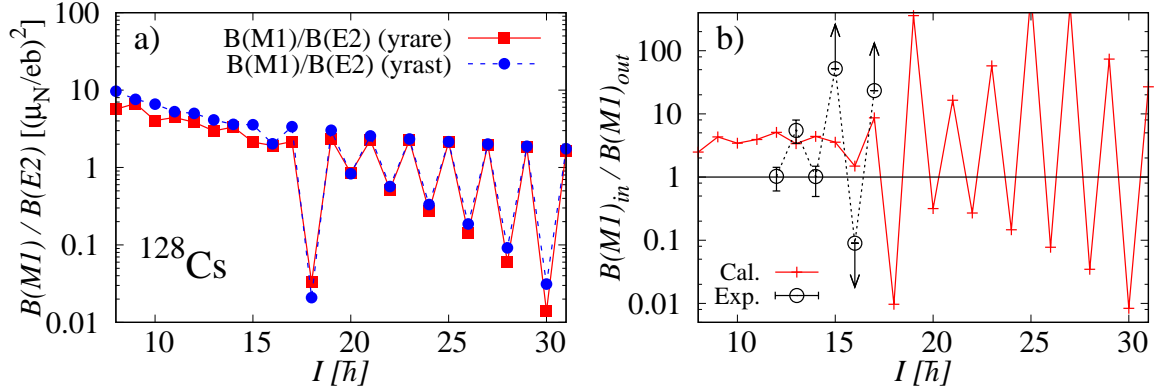}
\vspace*{-4mm}
\caption{(Color online)
Left panel: The calculated $B(M1:I\rightarrow I-1)/B(E2:I\rightarrow I-2)$
ratios inside the yrast band (solid lines)
and inside the yrare band (dashed lines) as functions of spin
for $^{128}$Cs.
Right panel: The ratio of the $I\rightarrow I-1$
in-band and out-of-band $M1$ transitions, $B(M1)_{\rm in}/B(M1)_{\rm out}$,
where the in-band transitions are those inside the yrare band
and the out-of-band transitions are those from the yrare to the yrast band.
As for $B(M1)_{\rm in}/B(M1)_{\rm out}$
the experimental data are also included~\cite{KSC03}.
These are the results of projection with the configuration-mixing
corresponding to Fig.~\ref{fig:sCsmixed}.
}
\label{fig:trCsmixed}
\end{center}
\end{figure*}

The characteristic geometry of the static chirality is
reflected in the ${\mit\Delta}I=1$ electromagnetic transition rates
as it is reviewed in Sec.~\ref{sec:selrule}.
We show the $I\rightarrow I-1$ $E2$ and $M1$ transition rates as functions
of spin in Fig.~\ref{fig:tEMCsmixed} (both in-band and out-of-band transitions).
It is clearly seen that the behavior of both $E2$ and $M1$ transitions
changes around $I=16$.
The $B(E2)$ and $B(M1)$ for the yrast and yrare bands become
similar after the chiral geometry is realized in $I \gtsim 16$.
For $I \ltsim 15$ the in-band transitions are larger than the out-of-band
transitions.  For $I \gtsim 16$ the in-band and out-of-band transitions
are of similar magnitude, and both of them
show the characteristic zigzag pattern.
Especially, the in-band (out-of-band) transition rates are prohibitively
small at even (odd) spins, and which transition is stronger,
the in-band or the out-of-band, changes alternatively as a function of spin.
This is exactly what is expected from the prototype model of Ref.~\cite{KSH04}
(see Fig.~\ref{fig:selchrl}).
As often shown in the experimental data,
the $B(M1)/B(E2)$ ratios for the yrast and yrare bands
are displayed in a logarithmic scale
in the left panel of Fig.~\ref{fig:trCsmixed}.
Again, the behavior of the ratios changes after the critical spin
and clearly shows a regular zigzag pattern,
which comes from the $M1$ transitions.
As it is discussed in Sec.~\ref{sec:selrule},
the clear signature of this ideal scenario can be seen
for the ratio of in-band versus out-of-band $M1$ transitions,
which is compared with experimental data~\cite{KSC03}
in the right panel of Fig.~\ref{fig:trCsmixed}.
For $I \gtsim 17$ this ratio alternates values greater than one at odd spin
and smaller than one at even spin alternatively,
which well corresponds to experimentally observed feature,
although the spin range is slightly shifted
as expected from the excitation energies
in the right panel of Fig.~\ref{fig:sCsmixed}.
In this way, the result of the present microscopic calculation
clearly shows the characteristic behavior of the chiral geometry
predicted by the phenomenological model of Ref.~\cite{KSH04}
not only for the energy spectrum
but also for the transition rates.

\begin{figure*}[!htb]
\begin{center}
\includegraphics[width=155mm]{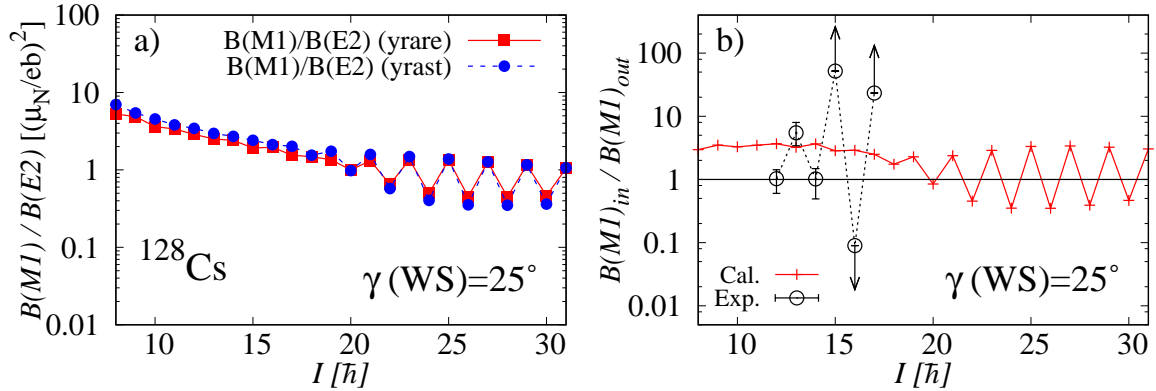}
\vspace*{-4mm}
\caption{(Color online)
The same as Fig.~\ref{fig:trCsmixed} but the calculation
with using $\gamma({\rm WS})=25^\circ$.
}
\label{fig:trCsmixedg25}
\end{center}
\end{figure*}

To see the effect of the triaxial deformation
we have done the same calculation using smaller values
of $\gamma$ than $30^\circ$:
The resultant energy difference between the yrast and yrare bands
increases and the amplitude of the zigzag behavior of
the $I\rightarrow I-1$ $E2$ and $M1$ transitions decreases,
while the $I\rightarrow I-2$ rotational $E2$ transitions
do not essentially change.  As an example, we show the result of
the $B(M1)/B(E2)$ and $B(M1)_{\rm in}/B(M1)_{\rm out}$ ratios
in Fig.~\ref{fig:trCsmixedg25}, which is obtained by the calculation
using $\gamma({\rm WS})=25^\circ$ and keeping the other parameters unchanged.
Note that $\gamma({\rm WS})=25^\circ$ corresponds to
$\gamma({\rm den})=19.3^\circ$.  As it is seen in Fig.~\ref{fig:momiXe},
the moment of inertia ${\cal J}_x$ is about factor two larger than ${\cal J}_z$
in contrast to the case of $\gamma({\rm WS})=30^\circ$
with ${\cal J}_x\approx {\cal J}_z$, which is
the necessary condition for the model of Ref.~\cite{KSH04}.
The minimum energy difference between the yrast and yrare bands
in this case is about 340 keV at $I=21$, which is about factor three
larger than that in the case of $\gamma({\rm WS})=30^\circ$.
As it is clearly seen by comparing Fig.~\ref{fig:trCsmixedg25} with
Fig.~\ref{fig:trCsmixed}, the amplitude of the zigzag behavior
is reduced by one to two orders of magnitude.
Therefore, as it is emphasized in Ref.~\cite{KSH04},
the $B(M1)_{\rm in}/B(M1)_{\rm out}$ ratio tells
how well the situation of the model is realized.
These results are consistent with those
of the particle-rotor model in Ref.~\cite{QZW09},
where the results of calculations with several different $\gamma$ values
are presented.

\subsection{Chiral doublet band in $^{104}$Rh}
\label{sec:chiralRh}

As another example of a chiral doublet band in an odd-odd nucleus,
we present the result of calculation for $^{104}$Rh,
where the odd neutron occupies the particle-like
negative parity orbit (mainly $h_{11/2}$)
and the odd proton occupies the hole-like
positive parity orbit (mainly $g_{9/2}$).
In this case the high-$j$ orbits of the odd neutron and proton are different.
The resultant rotational band has negative parity.
The calculational procedure is the same as for $^{128}$Cs.
The adopted deformation parameters are $\beta_2=0.25$, $\beta_4=0.0$,
and $\gamma=30^\circ$, for which we have found that
a chiral doublet band appears in the calculations.
Note that $\gamma=\gamma({\rm WS})=30^\circ$ corresponds to
$\gamma({\rm den})=24.9^\circ$ in this case.
The average pairing gaps, calculated selfconsistently,
are $\Delta_{\rm n}=0.95$ MeV
and $\Delta_{\rm p}=0.76$ MeV for neutrons and protons, respectively.
The adopted value, $\beta_2=0.25$,
is again larger than the commonly used value,
$\beta_2 \approx 0.18-0.23$, in the nuclear region around $^{104}$Rh.
For this nucleus we show only the result of projection from
the non-cranked mean-field state constructed by the configuration (a)
in Eq.~(\ref{eq:sig4class}) for simplicity;
other results are qualitatively similar.

\begin{figure}[!htb]
\begin{center}
\includegraphics[width=155mm]{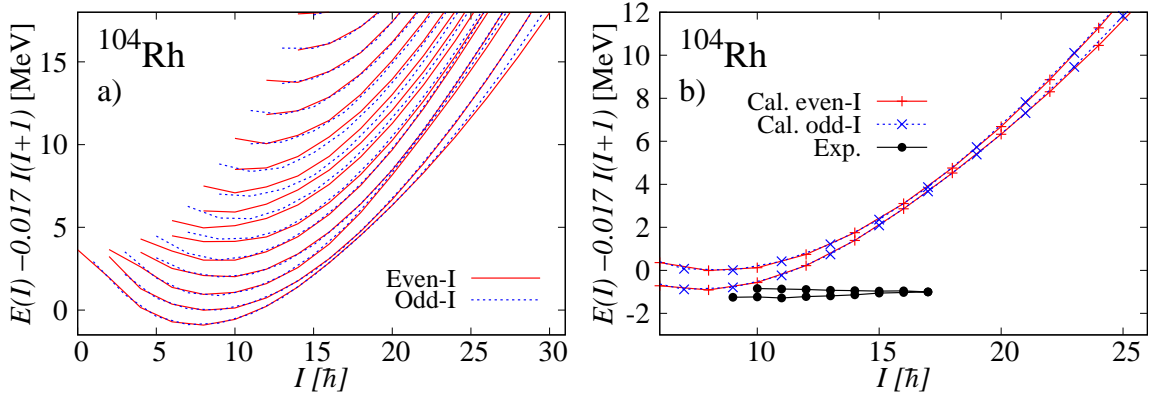}
\vspace*{-4mm}
\caption{(Color online)
Left panel: Energy spectrum for $^{104}$Rh calculated by
the angular-momentum-projection method
from the non-cranked mean-field
with the configuration (a) in Eq.~(\ref{eq:sig4class}).
A rigid-rotor reference energy of $0.017\,I(I+1)$ MeV is subtracted.
The lowest fourteen bands for both even-$I$ and odd-$I$ sequences are shown.
Right panel: Comparison of the calculated and experimental
chiral doublet bands in $^{104}$Rh.
Experimental data are taken from Ref.~\cite{VFK04}.
}
\label{fig:sRhnocr}
\end{center}
\end{figure}

\begin{figure*}[!htb]
\begin{center}
\includegraphics[width=155mm]{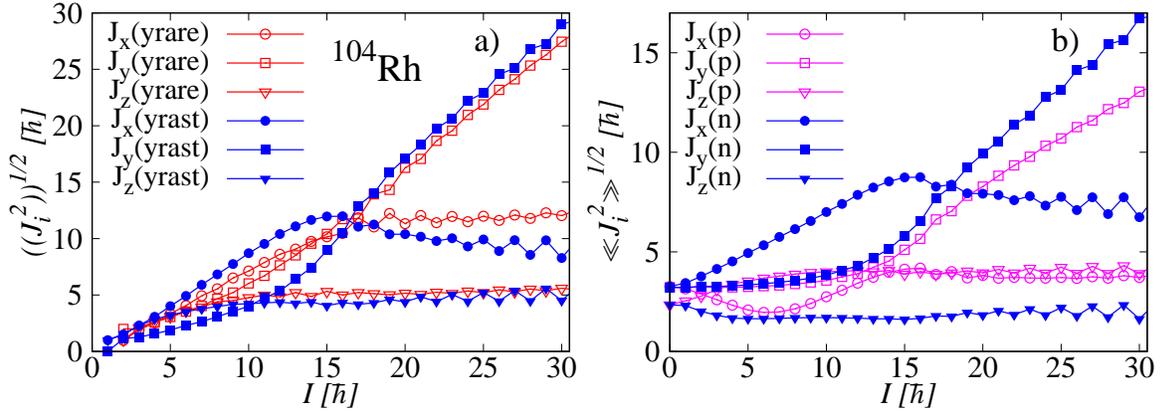}
\vspace*{-4mm}
\caption{(Color online)
The calculated expectation values of the angular-momentum vector
in the intrinsic frame for the non-cranked mean-field
of $^{104}$Rh corresponding to the spectrum in Fig.~\ref{fig:sRhnocr}.
The left panel shows the expectation values of the total vector
for the yrast (filled symbols) and yrare (open symbols)
${\mit\Delta}I=1$ bands, while the right panel shows
the neutron (filled symbols) and proton (open symbols) contributions
in Eq.~(\ref{eq:exJJm}) separately for the yrast band.
Note that the $x$, $y$, and $z$ axes are the short, medium, and long axes,
respectively.
}
\label{fig:jRhnocr}
\end{center}
\end{figure*}

The calculated spectrum is displayed in the left panel
of Fig.~\ref{fig:sRhnocr}.
The rigid-rotor reference energy $0.017\,I(I+1)$ MeV is
subtracted to see the details more clearly.
Just like the case of $^{128}$Cs in Fig.~\ref{fig:sCsnocr},
the even-$I$ and odd-$I$ sequences are nearly degenerate indicating
the signature symmetry is strongly broken.
The lowest two ${\mit\Delta}I=1$ bands, which are separated by
more than 1 MeV at low spins, quickly approach each other
within about 200$-$350 keV in the spin range $14 \ltsim I \ltsim 20$.
The estimated critical spin $I_{\rm c}\approx 12$ in Eq.~(\ref{eq:critI})
is slightly smaller than the spin
where the two bands become almost degenerate.
This behavior rather well corresponds to the observed one~\cite{VFK04},
although the moments of inertia of these bands are underestimated
compared with the experimental data
as it is shown in the right panel of Fig.~\ref{fig:sRhnocr}.
The chiral geometry is confirmed also in this case
by the expectation values of the angular-momentum vector in the intrinsic frame,
which are depicted in Fig.~\ref{fig:jRhnocr}.
At the low spins, $I \ltsim 8$,
the components of the short ($x$) and long ($z$) axes are dominant
for the yrast band, while the component of the medium ($y$) axis quickly grows.
All three components give important contributions
at the intermediate spin region;
see the left panel of Fig.~\ref{fig:jRhnocr}.
As it is discussed in the case of $^{128}$Cs,
the $y$ component of the angular-momentum vector for the yrare band
is considerably larger than that for the yrast band at $I \ltsim 15$,
which indicates that the yrare band can be interpreted as
one-phonon excitation of the chiral vibration at this lower spin region.
As expected, for $I \gtsim 16$ the behavior of the angular-momentum vectors
indicates that the system is in the regime of the static chirality.

Looking into the right panel of Fig.~\ref{fig:jRhnocr},
where the neutron and proton contributions are displayed separately,
the main contribution comes from the neutron for the $x$ component
and from the proton for the $z$ component,
while both neutron and proton coherently contribute to the $y$ component
as expected for collective angular-momentum.
The axis with the largest moment inertia is the $y$ axis as it is seen
from the cranking inertias of the core nucleus in Fig.~\ref{fig:momiPd}.
This collective angular-momentum component is very small below
$I_{\rm c}$ and starts to increase at higher spins $I>I_{\rm c}$.
Below $I_{\rm c}$, the collective part lies mainly in the $xz$ plane
with the $x$ component being more favored as in the case of $^{128}$Cs,
but neutrons contribute to it more than protons in contrast to $^{128}$Cs.
This behavior is again consistent with the model in Sec.~\ref{sec:selrule}.
Thus, the expected transition from the regime of the chiral vibration
to that of the static chirality is also confirmed in this case.

\begin{figure*}[!htb]
\begin{center}
\includegraphics[width=155mm]{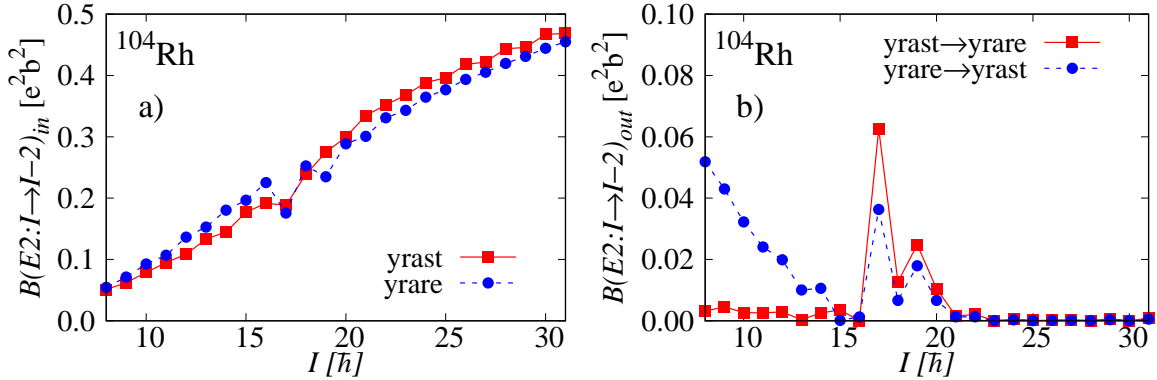}
\vspace*{-4mm}
\caption{(Color online)
The calculated $B(E2:I\rightarrow I-2)$ values for
the yrast band (solid lines) and the yrare band (dashed lines)
as functions of spin in $^{104}$Rh.
The left and right panels show the in-band and out-of-band transition rates,
respectively.  Note that the ordinate scale in the right panel is different
from that in the left panel.
These are the results of projection from the non-cranked lowest
configuration (a) in Eq.~(\ref{eq:sig4class})
corresponding to Fig.~\ref{fig:sRhnocr}.
}
\label{fig:tRRhnocr}
\end{center}
\end{figure*}

The $B(E2)$ and $B(M1)$ values inside the yrast and yrare bands
as well between the two bands are also calculated for $^{104}$Rh.
The $I\rightarrow I-2$ stretched $B(E2)$ values are displayed
in Fig.~\ref{fig:tRRhnocr}.
It is seen that the in-band values are large and
are similar for the yrast and the yrare bands.
These rotational transition probabilities increase as functions of spin,
which corresponds to the fact that the direction of
the angular-momentum vector changes gradually to the medium ($y$) axis
as the spin increases, as it is shown in Fig.~\ref{fig:jRhnocr}.
The out-of-band transitions are non-negligible for $16 \le I \le 19$,
where the difference of energies between the two bands is very small
and band mixing is expected.
These features are very similar to the case of $^{128}$Cs.
The $B(E2)$ values seem to be overestimated compared with
the experimental data~\cite{SRK08},
because the deformation parameter $\beta_2=0.25$ may be too large.

\begin{figure*}[!htb]
\begin{center}
\includegraphics[width=155mm]{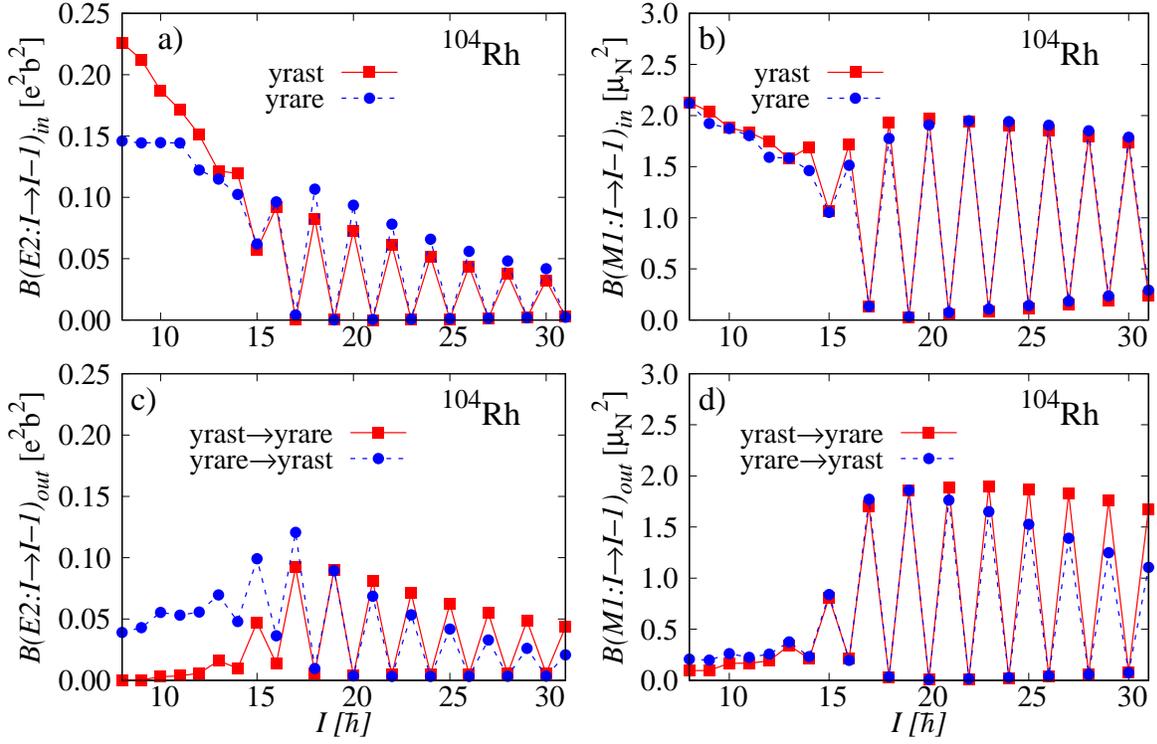}
\vspace*{-4mm}
\caption{(Color online)
The calculated
$B(E2:I\rightarrow I-1)$ and $B(M1:I\rightarrow I-1)$ values
for the yrast band (solid lines) and the yrare band (dashed lines)
as functions of spin in $^{104}$Rh.
The upper (lower) panels show the in-band (out-of-band) transition rates.
These are the results of projection from the non-cranked lowest
configuration (a) in Eq.~(\ref{eq:sig4class})
corresponding to Fig.~\ref{fig:sRhnocr}.
}
\label{fig:tEMRhnocr}
\end{center}
\end{figure*}

\begin{figure*}[!htb]
\begin{center}
\includegraphics[width=155mm]{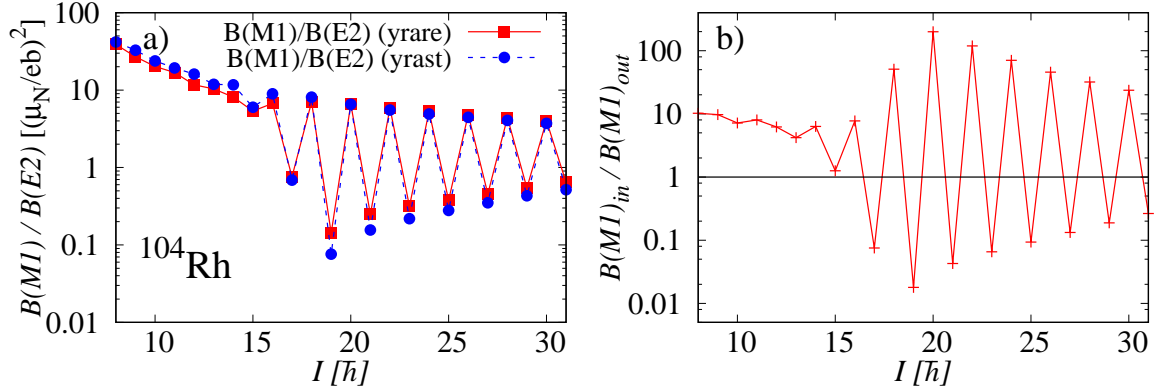}
\vspace*{-4mm}
\caption{(Color online)
Left panel: The calculated $B(M1:I\rightarrow I-1)/B(E2:I\rightarrow I-2)$
ratios inside the yrast band (solid lines)
and inside the yrare band (dashed lines) as functions of spin
for $^{104}$Rh.
Right panel: The ratio of the $I\rightarrow I-1$
in-band and out-of-band $M1$ transitions, $B(M1)_{\rm in}/B(M1)_{\rm out}$,
where the in-band transitions are those inside the yrare band
and the out-of-band transitions are those from the yrare to the yrast band.
These are the results of projection from the non-cranked mean-field
corresponding to Fig.~\ref{fig:sRhnocr}.
}
\label{fig:trRhnocr}
\end{center}
\end{figure*}

The calculated $B(E2:I\rightarrow I-1)$ and $B(M1:I\rightarrow I-1)$ values
are shown in Fig.~\ref{fig:tEMRhnocr}
in the same way as in the case of $^{128}$Cs.
The behavior of both $B(E2)$ and $B(M1)$ as functions of spin changes
at around $I\approx 15$; after this spin they exhibit typical zigzag behavior,
which is expected after the static chirality is realized.
As it is predicted in the prototype model of Ref.~\cite{KSH04},
the in-band transitions are large
when the out-of-band transitions are small and vice versa.
In case of $^{104}$Rh the in-band (out-of-band) transitions
are almost negligible for odd (even) spins,
which is opposite to the case of $^{128}$Cs.
This may be expected because both the particle and hole orbits are
mainly $h_{11/2}$ in $^{128}$Cs,
while the proton hole orbit in $^{104}$Rh is mainly $g_{9/2}$:
If one $h_{11/2}$ hole is replaced with $g_{g/2}$,
the coupled total spin may be reduced by one unit.
Finally, the $B(M1)/B(E2)$ ratio and
the in-band versus out-of-band $B(M1)$ ratio are shown in the logarithmic scale
in the left and right panels of Fig.~\ref{fig:trRhnocr}, respectively.
As discussed in the case of $^{128}$Cs,
The $B(M1)/B(E2)$ ratio changes to the regular zigzag behavior
after the chiral geometry is realized,
which reflects the behavior of the $B(M1)$ values.
The in-band versus out-of-band $B(M1)$ ratio also shows
a characteristic pattern, namely it alternates between values
greater than one and smaller than one as a function of spin.
This is the expected behavior from the model in Ref.~\cite{KSH04}
(see Fig.~\ref{fig:selchrl}).
However, it should be noted that the neutron-proton symmetry
prerequisite in the model of Ref.~\cite{KSH04}
is not precisely satisfied in the present example,
because the high-$j$ orbits of the odd neutron and proton are different.
It is interesting that the calculation shows
the characteristic selection rules of the model even in this case.
In fact, the particle-rotor model calculation with proton $g_{9/2}$
and neutron $h_{11/2}$ orbits in Ref.~\cite{JJR04} shows similar
zigzag behavior for $B(M1)$ for the neighboring nucleus $^{106}$Rh.
Although the zigzag behavior of $B(M1)$ are observed
in the experimental data~\cite{VFK04,SRK08},
its amplitude is too large in the present calculation.
The observed doublet band may not come as close to
the model of Ref.~\cite{KSH04} as our calculations.

\begin{figure*}[!htb]
\begin{center}
\includegraphics[width=155mm]{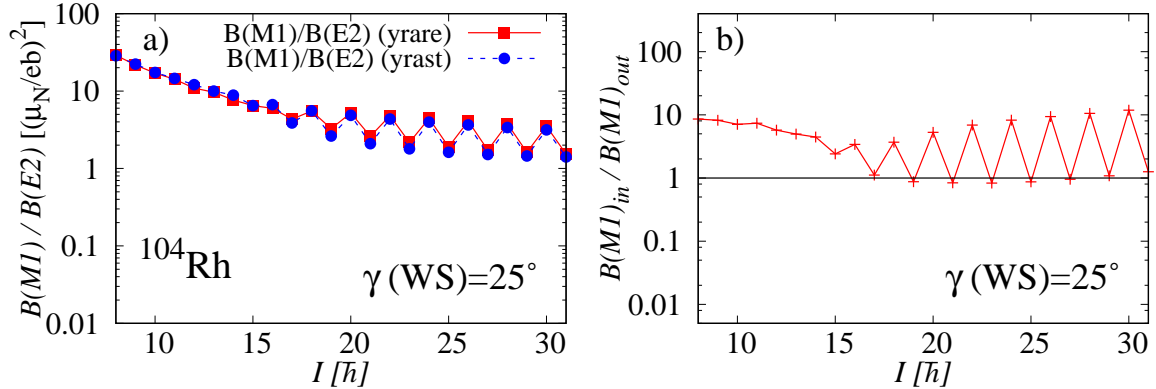}
\vspace*{-4mm}
\caption{(Color online)
The same as Fig.~\ref{fig:trRhnocr} but the calculation
with using $\gamma({\rm WS})=25^\circ$.
}
\label{fig:trRhnocrg25}
\end{center}
\end{figure*}

Finally we show the results of a calculation using
$\gamma=\gamma({\rm WS})=25^\circ$ for this nucleus.
Note that $\gamma=\gamma({\rm WS})=25^\circ$ corresponds to
$\gamma({\rm den})=20.2^\circ$ in this case.
The moment of inertia ${\cal J}_x$ is about factor two larger
than ${\cal J}_z$ (see Fig.~\ref{fig:momiPd}).
Figure~\ref{fig:trRhnocrg25} depicts
the $B(M1)/B(E2)$ and $B(M1)_{\rm in}/B(M1)_{\rm out}$ ratios
like Fig.~\ref{fig:trRhnocr}.  Apparently the magnitude of the oscillation
of the $B(M1)$ values are reduced by one to two orders of magnitude.
The out-of-band $B(M1)$ values become smaller
than the in-band values, and the center of the oscillations
is changed to a value that is considerably larger than one.

\section{Summary}
\label{sec:summary}

In this series of investigation, we have studied
rotational motion that is characteristic
for nuclei with triaxial deformation.
The basic method we employed is the fully microscopic framework
of angular-momentum projection from the mean-field wave function,
where the microscopic Hamiltonian is composed of
the Woods-Saxon mean-field and the separable schematic interaction.
Among various interesting types of rotational motion,
we have concentrated on
the nuclear wobbling motion and the chiral vibrations and rotations.
The former is the subject of part~I and
the latter is the subject of the present part~II in the series.

The nuclear chirality of rotating triaxially deformed nucleus
is a relatively new concept and it is expected in odd-odd nuclei
as typical examples.  We have applied our microscopic framework
to the typical cases of two odd-odd nuclei, $^{128}$Cs and $^{104}$Rh,
where the odd proton (neutron) occupies
the high-$j$ particle-like orbit and the odd neutron (proton) occupies
the high-$j$ hole-like orbit in the former (latter) nucleus.
The odd nucleons occupying the particle- and hole-like orbits align
their angular-momentum vectors along the short and long axes, respectively.
Combined with the collective rotation around the medium axis,
which has the largest moment of inertia,
these three angular-momentum vectors form an aplanar configuration,
i.e., the chiral geometry is realized in the body-fixed frame
of the triaxial mean-field.
In such a situation the chiral symmetry between the right- and left-handedness
is broken, which is the reason why the chiral doublet band emerges~\cite{FM97}.
Adjusting the quadrupole deformation parameter $\beta_2$ and
fixing the triaxiality parameter at $\gamma({\rm WS})=30^\circ$
we are able to obtain the yrast and yrare bands as a chiral doublet
by our fully microscopic angular-momentum-projection calculation.
By calculating the expectation values of the angular-momentum vector
with respect to the three principal axes,
it is confirmed that the chiral geometry is realized
for the selected examples of $^{128}$Cs and $^{104}$Rh.
However, the moments of inertia of the calculated bands are too small
compared with the experimental data.
One of the merits of the angular-momentum-projection method is
feasibility for calculating the electromagnetic transition probabilities.
We have studied the $E2$ and $M1$ transitions
between the members of the doublet bands.
It is demonstrated that the $I\rightarrow I-1$ transition rates
completely change their behavior after the static chirality is reached.
Large and small reduced probabilities alternate as functions of spin,
and this behavior is out of phase
for the in-band and out-of-band transitions.
This characteristic feature is in accordance with
the prototype model proposed in Ref.~\cite{KSH04},
and qualitatively corresponds to the experimental data
for both $^{128}$Cs and $^{104}$Rh.

In this way, we have confirmed that the two interesting types
of rotational motion, the wobbling motion and the chiral rotation,
which are characteristic for the triaxially deformed nuclei,
naturally emerge as results of
our fully microscopic angular-momentum-projection calculation.
The wobbling bands and chiral doublet bands were originally predicted
based on the macroscopic rotor model or
the phenomenological particle-rotor coupling model.
Considering the fact that the predicted properties by these models
are confirmed by our microscopic calculations,
the macroscopic rotor model picture is well realized
for the triaxially deformed nucleus.
It should, however, be noticed that a quantitative description
of these rotational bands is not achieved in the present series of work.
Further investigation is needed for a quantitative description of the data.


\vspace*{10mm}



\begin{thebibliography}{99}

\bibitem{BM75}
A.~Bohr and B.~R.~Mottelson,
{\it Nuclear Structure}, Vol.~II, Benjamin, New York (1975).

\bibitem{Mol06}
P.~M\"oller, R.~Bengtsson, B.~G.~Carlsson, P.~Olivius, and T.~Ichikawa
Phys.\ Rev.\ Lett.\ \textbf{97}, 162502 (2006).

\bibitem{VDS83}
M.~J.~A.~de Voigt,J.~Dudek and Z.~Szymanski,
Rev.\ Mod.\ Phys.\ \textbf{55}, 949 (1983).

\bibitem{Fra01}
S.~Frauendorf, Rev.\ Mod.\ Phys.\ \textbf{73}, 463 (2001).

\bibitem{Pan11}
S.~C.~Pancholi, "Exotic Nuclear Excitations",
Springer Tracts, in Modern Physics 242, Springer (2011).

\bibitem{Ode01}
S.~W.~\O{}deg\aa{}rd {\it et al.},
Phys.\ Rev.\ Lett.\ \textbf{86}, 5866 (2001).

\bibitem{NMM16}
T.~Nakatsukasa, K.~Matsuyanagi, M.~Matsuzaki, and Y.~R.~Shimizu,
Phys. Scripta {\bf 91}, 073008 (2016).

\bibitem{Fra17}
S.~Frauendorf,
Phys.\ Scripta, to be published, arXiv:1710.01210.

\bibitem{SFTSI}
M.~Shimada, Y.~Fujioka, S.~Tagami, and Y.~R.~Shimizu,
to be published.

\bibitem{FM97}
S.~Frauendorf and J.~Meng,
Nucl.\ Phys.\ A \textbf{617}, 131 (1997).

\bibitem{SK17}
K.~Starosta and T.~Koike,
Phys.\ Scripta {\bf 92}, 093002 (2017).

\bibitem{KSH04}
T.~Koike, K.~Starosta, and I.~Hamamoto,
Phys.\ Rev.\ Lett.\ \textbf{93}, 172502 (2004).

\bibitem{RS80}
P.~Ring and P.~Schuck,
{\it The nuclear many-body problem}, Springer (1980).

\bibitem{TS12}
S.~Tagami and Y.~R.~Shimizu,
Prog. Theor. Phys. \textbf{127}, 79 (2012).

\bibitem{TSD13}
S.~Tagami, Y.~R.~Shimizu, and J.~Dudek,
Phys.\ Rev.\ C \textbf{87}, 054306 (2013).

\bibitem{TSD15}
S.~Tagami, Y.~R.~Shimizu, and J.~Dudek,
J.\ Phys.\ G {\bf 42}, 015106 (2015).

\bibitem{TS16} 
S.~Tagami and Y.~R.~Shimizu,
Phys.\ Rev.\ C {\bf 93}, 024323 (2016).

\bibitem{STS15}
M.~Shimada, S.~Tagami, and Y.~R.~Shimizu,
Prog. Theor. Exp. Phys. \textbf{2015}, 063D02 (2015).

\bibitem{STS16}
M.~Shimada, S.~Tagami, and Y.~R.~Shimizu,
Phys.\ Rev.\ C \textbf{93}, 044317 (2016).

\bibitem{BSP12}
G.~H.~Bhat, J.~A.~Sheikh, and R.~Palit,
Phys.\ Lett.\ B \textbf{707}, 250 (2012).

\bibitem{Sun16}
Y.~Sun, Phys.\ Scripta \textbf{91}, 043005 (2016).

\bibitem{SBD16}
J.~A.~Sheikh, G.~H.~Bhat, W.~A.~Dar, S.~Jehangir, and P.~A.~Ganai,
Phys.\ Scripta {\bf 91}, 063015 (2016).

\bibitem{TSF14}
S.~Tagami, M.~Shimada, Y.~Fujioka, Y.~R.~Shimizu, and J.~Dudek,
Physica Scripta \textbf{89}, 054013 (2014).

\bibitem{BR85}
T.Bengtsson and I.~Ragnarsson,
Nucl.\ Phys.\ A \textbf{436}, 14 (1985).

\bibitem{SSM08}
Y.~R.~Shimizu, T.~Shoji and M.~Matsuzaki,
Phys. Rev. C {\bf 77}, 024319 (2008).

\bibitem{SKC01}
K.~Starosta et al.,
Phys.\ Rev.\ Lett.\ \textbf{86}, 971 (2001).

\bibitem{KSC03}
T.~Koike et al.,
Phys.\ Rev.\ C \textbf{67}, 044319 (2003).

\bibitem{ADF11}
D.~Almehed, F.~D\"onau, and S.~Frauendorf,
Phys.\ Rev.\ C \textbf{83}, 054308 (2011).

\bibitem{ODD04}
P.~Olbratowski, J.~Dobaczewski, J.~Dudek, and W.~P{\l}{\'o}ciennik,
Phys.\ Rev.\ Lett.\ \textbf{93}, 052501 (2004).

\bibitem{VFD00}
V.~I.~Dimitrov, S.~Frauendorf, and F.~D\"onau,
Phys.\ Rev.\ Lett.\ \textbf{84}, 5732 (2000).

\bibitem{ZGN03}
S.~Zhu et al.,
Phys.\ Rev.\ Lett.\ \textbf{91}, 132501 (2003).

\bibitem{QZM09}
B.~Qi, S.~Q.~Zhang, J.~Meng, S.~Y.~Wang, and S.~Frauendorf,
Phys.\ Lett.\ B \textbf{675}, 175 (2009).

\bibitem{QZW09}
B.~Qi, S.~Q.~Zhang, S.~Y.~Wang, J.~M.~Yao, and J.~Meng,
Phys.\ Rev.\ C\ \textbf{79}, 041302(R) (2009).

\bibitem{Fra93}
S.~Frauendorf, Nucl.\ Phy.\ \textbf{557}, 259c (1993).

\bibitem{Fra00}
S.~Frauendorf, Nucl.\ Phy.\ \textbf{677}, 115 (2000).

\bibitem{GSP06}
E.~Grodner et al.,
Phys.\ Rev.\ Lett.\ \textbf{97}, 172501 (2006).

\bibitem{VFK04}
C.~Vaman et al.,
Phys.\ Rev.\ Lett.\ \textbf{92}, 032501 (2004).

\bibitem{SRK08}
T.~Suzuki et al.,
Phys.\ Rev.\ C\ \textbf{78}, 031302(R) (2008).

\bibitem{JJR04}
P.~Joshi et al.,
Phys.\ Lett.\ B\ \textbf{595}, 135 (2004).


\end{thebibliography}
\end{document}